\documentclass[final,7p, twocolumn]{elsarticle}

\usepackage{amssymb}
\usepackage{amsmath}
\usepackage{cleveref}
\usepackage{todonotes}
\usepackage{balance}
\usepackage{enumerate}
\usepackage{url}
\usepackage[T1]{fontenc}

\usepackage{standalone}
\usepackage{graphicx}
\usepackage{tabularx}
\usepackage{booktabs}
\usepackage[table]{xcolor}
\usepackage{multirow} 
\usepackage{color, colortbl}

\usepackage{amsthm}

\usepackage{nicefrac}

\usepackage{cclicenses}

\usepackage{caption}
\usepackage{subcaption}
\usepackage[ruled,vlined,linesnumbered]{algorithm2e}
\usepackage{algpseudocode}

\usepackage{todonotes}

\usepackage{textcomp}
\usepackage{verbatim}
\usepackage{listings}
\usepackage{mdframed}

\definecolor{backcolour}{rgb}{0.9,0.9,0.9}

\lstdefinestyle{mystyle}{
    backgroundcolor=\color{backcolour}
}

\usepackage{amsthm}
\theoremstyle{definition}
\newtheorem{definition}{Definition}

\lstdefinelanguage{yaml}{
  keywords={true, false, null, yes, no},
  sensitive=true,     
  morecomment=[l]{\#},
  morestring=[b]',    
  morestring=[b]",    
  morestring=[b]|,    
  morestring=[b]>,    
  literate=           
    {:}{{{\color{red}:}}}{1}
    {,}{{{\color{red},}}}{1}
    {?}{{{\color{red}?}}}{1}
    {!}{{{\color{red}!}}}{1}
    {>}{{{\color{red}>}}}{1}
    {|}{{{\color{red}|}}}{1}
    {\ }{{\ }}{1},
  columns=fullflexible,
}

\newenvironment{yamlcodep}[1]
  {\noindent\begin{minipage}{#1}
  \begin{mdframed}[linecolor=white,backgroundcolor=yellow!25]}
  {\end{mdframed}\end{minipage}}

\journal{Future Generation Computer Systems}

\begin{document}

\begin{frontmatter}

%% Title, authors and addresses

%% use the tnoteref command within \title for footnotes;
%% use the tnotetext command for theassociated footnote;
%% use the fnref command within \author or \affiliation for footnotes;
%% use the fntext command for theassociated footnote;
%% use the corref command within \author for corresponding author footnotes;
%% use the cortext command for theassociated footnote;
%% use the ead command for the email address,
%% and the form \ead[url] for the home page:
%% \title{Title\tnoteref{label1}}
%% \tnotetext[label1]{}
%% \author{Name\corref{cor1}\fnref{label2}}
%% \ead{email address}
%% \ead[url]{home page}
%% \fntext[label2]{}
%% \cortext[cor1]{}
%% \affiliation{organization={},
%%             addressline={},
%%             city={},
%%             postcode={},
%%             state={},
%%             country={}}
%% \fntext[label3]{}

\title{Green by Design: Constraint-Based Adaptive Deployment in the Cloud Continuum} %% Article title

%% use optional labels to link authors explicitly to addresses:
%% \author[label1,label2]{}
%% \affiliation[label1]{organization={},
%%             addressline={},
%%             city={},
%%             postcode={},
%%             state={},
%%             country={}}
%%
%% \affiliation[label2]{organization={},
%%             addressline={},
%%             city={},
%%             postcode={},
%%             state={},
%%             country={}}

%\author{} %% Author name
\author[label1]{Andrea D'Iapico}
\author[label1]{Monica Vitali}\corref{monica.vitali@polimi.it}

%% Author affiliation
\affiliation[label1]{organization={Politecnico di Milano},
            addressline={Piazza Leonardo 32}, 
            city={Milan},
            postcode={20133}, 
            country={Italy}}
\begin{abstract}

The environmental sustainability of Information Technology (IT) has emerged as a critical concern, driven by the need to reduce both energy consumption and greenhouse gas (GHG) emissions. In the context of cloud-native applications deployed across the cloud–edge continuum, this challenge translates into identifying energy-efficient deployment strategies that consider not only the computational demands of application components but also the environmental impact of the nodes on which they are executed.
Generating deployment plans that account for these dynamic factors is non-trivial, due to fluctuations in application behaviour and variations in the carbon intensity of infrastructure nodes. In this paper, we present an approach for the automatic generation of deployment plans guided by green constraints. These constraints are derived from a continuous analysis of energy consumption patterns, inter-component communication, and the environmental characteristics of the underlying infrastructure.

This paper introduces a methodology and architecture for the generation of a set of green-aware constraints that inform the scheduler to produce environmentally friendly deployment plans. We demonstrate how these constraints can be automatically learned and updated over time using monitoring data, enabling adaptive, energy-aware orchestration. The proposed approach is validated through realistic deployment scenarios of a cloud-native application, showcasing its effectiveness in reducing energy usage and associated emissions.
\end{abstract}

% %%Graphical abstract
% \begin{graphicalabstract}
% %\includegraphics{grabs}
% \end{graphicalabstract}

% %%Research highlights
% \begin{highlights}
% \item Research highlight 1
% \item Research highlight 2
% \end{highlights}

%% Keywords
\begin{keyword}
Green IT \sep Cloud Native Applications \sep Cloud Computing \sep Cloud Continuum \sep Deployment Plan
%% keywords here, in the form: keyword \sep keyword

%% PACS codes here, in the form: \PACS code \sep code

%% MSC codes here, in the form: \MSC code \sep code
%% or \MSC[2008] code \sep code (2000 is the default)

\end{keyword}

\end{frontmatter}

\section{Introduction} \label{sec:intro}
Data centres are among the primary contributors to the environmental impact of the Information Technology (IT) sector. Recent initiatives by cloud providers have significantly improved the infrastructural efficiency of data centres, with Power Usage Effectiveness (PUE) values approaching the ideal of 1.0. The adoption of renewable energy sources to power large-scale data centres has further reduced their environmental footprint. 
However, the rapid and continuous growth in the demand for computational resources has shifted the sustainability focus from the infrastructure level to the application level. At the same time, the emergence of edge computing and the widespread adoption of the cloud continuum for service deployment have led to a highly heterogeneous environment, where many computational nodes fall short of the efficiency standards of centralised data centres, while satisfying security and privacy constraints.
In this context, DevOps engineers can play a pivotal role by designing microservice-based applications that are capable of adapting their composition based on the energy consumption profiles of individual services and the available energy or carbon budget, and by orchestrating energy-aware deployments across heterogeneous infrastructures. To maintain high QoS while optimising sustainability, deployment decisions must be adaptable and context-aware. Defining green-aware deployment constraints is a complex task due to the dynamic nature of the system and the difficulty of predicting the actual environmental impact of specific design decisions.

Our solution assists DevOps engineers in deploying environmentally sustainable applications by automatically generating deployment constraints that can reduce the energy consumption and lower the environmental footprint of service-based applications. We define and classify constraint types that support greener deployment strategies, and propose a methodology to automatically generate these constraints, assigning them dynamic, context-specific relevance through continuously updated weights.
The approach is validated with a benchmark microservice-based application, demonstrating the adaptability of the green constraints in response to evolving contextual conditions, and highlighting their potential to reduce environmental impact. The main contributions of this paper are:
\begin{itemize}
    \item the definition and classification of deployment constraints tailored for green-by-design microservice-based applications;
    \item a methodology and architecture for the automatic generation and continuous refinement of these constraints, based on monitoring data and infrastructure characteristics;
    \item the creation of an \textit{Explainability Report}, offering human-readable insights to support DevOps engineers in making deployment decisions.
\end{itemize}

The remainder of the paper is structured as follows. In Sect.~\ref{sec:related}, we review the state-of-the-art. Section~\ref{sec:methodology} introduces the proposed methodology and architecture, while in Sect.~\ref{sec:constraints}, we detail the modelling and implementation of the proposed solution. Section~\ref{sec:validation} presents a case study that demonstrates the effectiveness of our approach. Finally, Sect.~\ref{sec:conclusion} concludes the paper and outlines directions for future work.

\section{State of the Art} \label{sec:related}
The environmental impact of Information Technology (IT) has been a significant concern in recent years, due to the significant impact of the operations executed in the big cloud data centres and the amount of data transmitted daily on the internet. To limit this impact, cloud providers are promoting actions towards carbon neutrality. These actions also aim to increase the transparency in the carbon reporting and the availability of carbon tracking tools~\cite{GoogleCloud2023CarbonFootprint,Microsoft2023EmissionsImpact,AWS2023CarbonFootprintTool,meta2024retinas}.
However, the increasing demand for computational and storage resources in the cloud~\cite{snp2024datacenter_trends}, coupled with the slow transition toward renewable energy sources, has limited the positive outcomes of these initiatives~\cite{Lawrence2024NetZero}, and the trend in the greenhouse gas (GHG) emissions in data centres is still increasing~\cite{Google2024EnvReport,Microsoft2024EnvReport,meta2023sustainability}.

As a consequence, the sole effort of the infrastructure providers is not sufficient to reach carbon neutrality in IT operations. An effort is also necessary in the design, orchestration, and management of the applications executed in those infrastructures. This family of approaches goes under the umbrella of carbon-aware computing, a key paradigm for improving energy efficiency of the execution of applications while measuring and reducing the carbon footprint. Carbon Intensity (CI) is a metric that refers to the amount of GHG emissions generated per unit of electricity consumed, expressed in grams of CO$_2$ per kilowatt-hour (gCO$_2$/kWh). This value is dependent on the location in which the application is executed (different regions have different energy mixes) as well as the time of execution (due to the intermittent availability of renewable resources). Carbon-aware applications are designed to adapt their execution based on the current energy mix available. 

Energy-efficient deployment and execution of cloud applications have been studied with promising results. In \cite{de2021revisiting}, a comparison in terms of efficiency between monolithic and microservices architectures is performed, highlighting the higher efficiency of the latter and proposing a dynamic resource provisioning approach to reduce energy consumption while maintaining performance. In~\cite{10500420}, authors identify the aspects that mainly influence the energy efficiency of microservice-based applications. Modern infrastructures combine a heterogeneous set of resources, geographically distributed with different efficiency and computational capabilities, such as cloud-edge infrastructures. In this scenario, workflow scheduling is more challenging. In \cite{zhang2017bi}, the authors present a scheduling approach for cloud-edge infrastructures, balancing energy efficiency and system reliability.

Traditionally, carbon-aware computing techniques focus on two main approaches: location shifting, where the workload is moved to a more convenient region, and time shifting, where batch operations are postponed to a moment where a lower carbon intensity is available. Several studies have proposed this type of approaches~\cite{wiesner2021letswaitawhile,Radovanovic_Google_2022,lechowicz2023online,Zheng_MitigatingCurtailment_2020,wiesner2024fedzero,hanafy2024asplos,sukprasert2024eurosys}, with the limitation to fit only batch operations. 
For example, \cite{choudhary2022energy} proposes a power-aware scheduling algorithm that ensures task completion within user-defined deadlines while reducing the energy consumption of scientific workflows. 

However, time-sensitive cloud applications requiring low-latency interactions present different challenges. The main approaches available for this type of application involve geographical load balancing and capacity scaling~\cite{Zhou2013Carbon-Aware,Zhou2016Carbon-Aware}. Several approaches have demonstrated how it is possible to reduce the emissions of time-sensitive applications while maintaining their performance~\cite{souza2023casper,murillo2024cdn_shifter,gsteiger2024caribou}. In \cite{10592774}, authors use a genetic algorithm to change the deployment of microservice-based applications considering the impact of sustainability on performance, while a reinforcement learning approach is applied in~\cite{10680533}, and a particle swarm optimisation is proposed in~\cite{10683711}.
However, geographical distribution is not always a feasible solution, due to constraints related to latency, data regulations, privacy, and infrastructure limitations. Moreover, these techniques alone do not guarantee that a feasible configuration meeting energy or budget constraints always exists.

Complementary approaches have been recently proposed to enhance the set of possible adaptive actions that can be employed to reduce the environmental impact of microservice-based applications dynamically, exploiting approximation techniques~\cite{barua2019approximate,leon2025approximate}. %In machine learning tasks, approximation can improve energy efficiency by reducing computational precision~\cite{zervakis2021approximate,10682772}, also in edge-based ML tasks \cite{irtija2021energy}, or focusing on training dataset reduction~\cite{anselmo2023data}.
In workflow management, approximation involves using alternative implementations or execution modalities of functionalities to optimise energy efficiency, favouring certain non-functional requirements over others. In \cite{stavrinides2019energy}, energy consumption is reduced through Dynamic Voltage and Frequency Scaling (DVFS), trading off precision while maintaining acceptable result quality. In cloud computing, this idea aligns with \textit{graceful degradation}, where the application's features are dynamically disabled in response to overload conditions~\cite{meta2023defcon}.

The exploitation of approximation features is possible only if such features are explicitly expressed in the application design. The Sustainable Application Design Process (SADP)~\cite{vitali2022towards,vitali2023enriching} defines approximation features and incorporates them through metadata into application design. SADP supports the definition of multiple flavours of the same functionality with different quality and energy trade-offs. SADP also allows the declaration of optional services in the application workflows, disabled in case of high energy consumption. Such features enable microservice re-orchestration, but generate potential performance degradation~\cite{zambianco2024cost}, necessitating careful selection of re-orchestration intervals. The automation in the selection of the best configuration for a specific infrastructure at a specific time is not provided in SADP. A scheduler able to support these features is needed, as well as mechanisms to inform the scheduler on the best deployment choices for a specific context.

We present an evolution of this approach, where energy-efficiency constraints for the workflow deployment are automatically generated. The approach is part of the FREEDA (Failure-Resilient, Energy-aware, and Explainable Deployment of microservice-based Applications over Cloud-IoT infrastructures) project, supporting the generation of failure-resilient and energy-aware deployment plans, exploiting approximation features expressed as optional services as well as multiple services flavours~\cite{gazza2025constraint}.
%~\cite{amadini2024pick,vitali2024freeda,soldani2024towards,gazza2025constraint}.

\section{Methodology and Architecture}\label{sec:methodology}

\begin{figure*}[t]
    \centering
    \includegraphics[width=.9\textwidth]{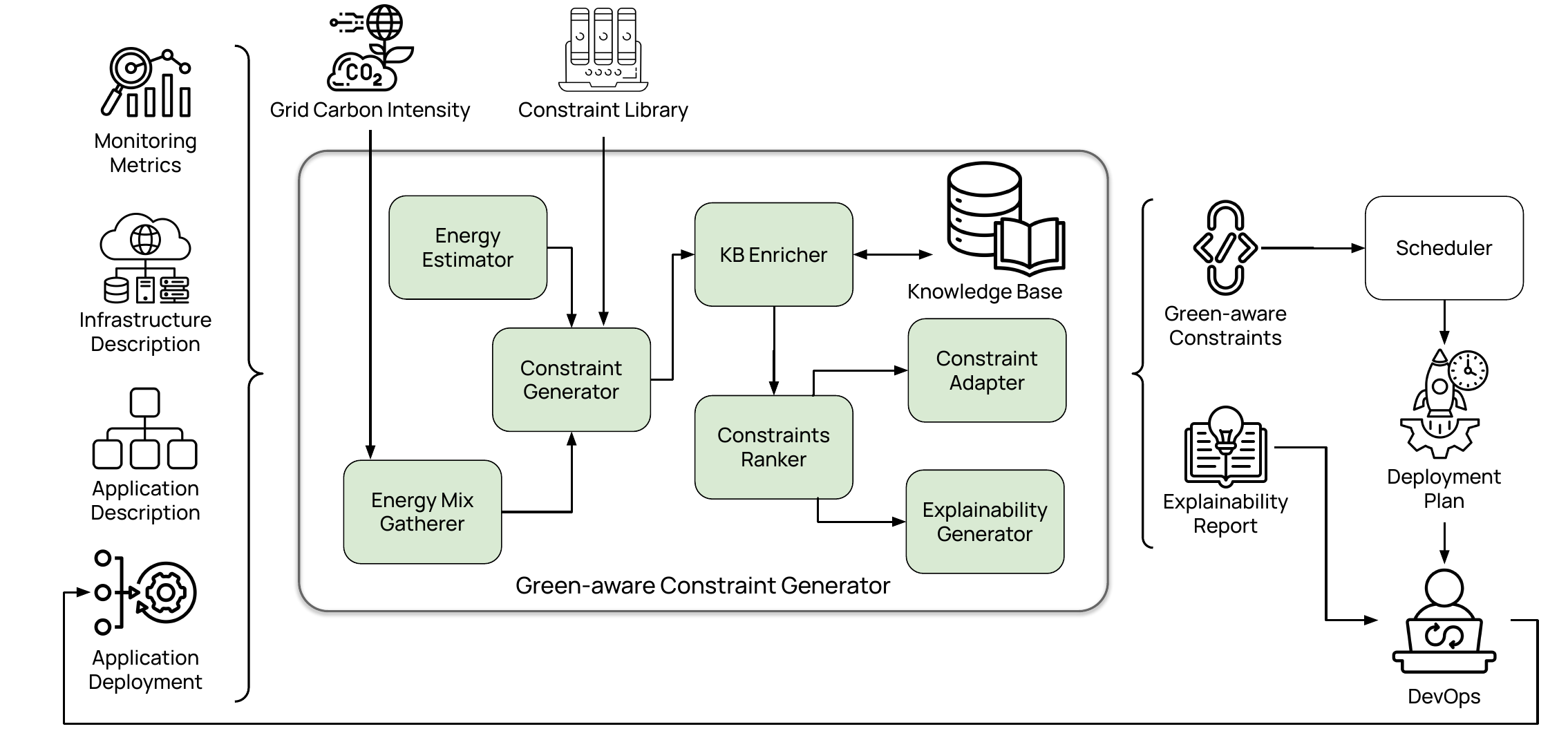}
    \caption{Overview of the Green-aware Constraints Generation methodology.}
    \label{fig:architecture}
\end{figure*}

Deploying composite applications across distributed and heterogeneous infrastructures is a complex task. 
Modern schedulers can generate feasible and efficient deployment plans; however, current state-of-the-art solutions overlook environmental considerations in their decision-making processes.

In this paper, we present a methodology to enable greener deployments of cloud-native applications within the cloud continuum. 
Our approach introduces a \textit{Green-aware Constraint Generator}, which takes as input the \textit{Application Description} (i.e., the services to be deployed), the \textit{Infrastructure Description} (i.e., available nodes), the corresponding \textit{Grid Carbon Intensity}, and \textit{Monitoring Metrics} on application behaviour. 
From these inputs, it produces \textit{Green-aware Constraints} that guide a \textit{Scheduler} in generating a \textit{Deployment Plan}. 
This plan is then reviewed by the \textit{DevOps} engineer, who makes the final decision on the actual \textit{Application Deployment}. 
The methodology follows a Human-In-The-Loop (HITL) paradigm, in which the DevOps role is enhanced through an \textit{Explainability Report} that justifies the scheduler's recommended deployment choices. 
Furthermore, the approach supports adaptivity by iteratively learning and refining constraints over time. 
An overview of the methodology is shown in Fig.~\ref{fig:architecture}.

In our previous work, we demonstrated the feasibility of a scheduler capable of incorporating such constraints to produce valid deployment plans~\cite{gazza2025constraint}. 
Here, we focus on how the \textit{Green-aware Constraints} are generated, presenting in detail the architecture of the \textit{Green-aware Constraint Generator}.

The solution exhibits three key properties: (i) \textbf{Adaptivity}: the system automatically responds to changes in both the application and infrastructure by learning new constraints, while preserving and refining knowledge acquired in previous iterations; (ii) \textbf{Extensibility}: the types of constraints are defined within a dedicated \textit{Constraint Library}, which can be easily extended with additional constraint types with minimal effort; (iii) \textbf{Generality}: the approach is technology-agnostic; inputs are described using standard languages, and the resulting constraints can be adapted to dialogue with different schedulers.

\subsection{Green-Aware Constraint Generator Architecture}
The architecture of the \textit{Green-Aware Constraint Generator} is shown in Fig.~\ref{fig:architecture}, with the relevant modules coloured in green.

The \texttt{Energy Mix Gatherer} is responsible for enriching the \texttt{Infrastructure Description} with carbon intensity data for each node. This information is typically retrieved from publicly available services; hence, we assume the component accesses it through a \texttt{Grid Carbon Intensity} service. Since deployment decisions are not made instantaneously, the module considers the average carbon intensity over a recent observation window to provide a more stable and representative value.

The \texttt{Energy Estimator} focuses on estimating the energy consumption of both the application services and their intercommunications. Based on data collected from the \texttt{Monitoring Metrics}, it enriches the \texttt{Application Description} by associating energy profiles with each service and estimating the energy generated by data exchanges among them.

Using the enriched application and infrastructure descriptions as input, the \texttt{Constraint Generator} produces relevant green-aware deployment constraints. These constraints are defined based on the templates and rules stored in the \texttt{Constraint Library}. While this paper considers only two types of constraints, the library is modular and extensible, allowing additional constraint types to be easily incorporated. The constraints generated at this stage reflect the current behaviour within the current infrastructure, ensuring that the system remains up to date and adaptive to new configurations.

To maintain continuity across deployments, the \texttt{KB Enricher} retrieves previously learned constraints from an internal \texttt{Knowledge Base}, and updates it with newly generated constraints and refined energy profiles, ensuring that past knowledge is preserved and augmented over time.

Given that the set of past and present constraints can grow significantly, the \texttt{Constraint Ranker} component assigns an importance score to each constraint based on its predicted impact on energy efficiency and carbon emissions. Constraints with low expected impact are discarded, ensuring that only the most relevant ones are retained for consideration.

For the remaining high-impact constraints, the \texttt{Explainability Generator} produces an \textit{Explainability Report}. This report provides a rationale for each constraint, describing how it was generated and its estimated environmental impact. Finally, the \texttt{Constraint Adapter} reformats the constraints into a syntax and structure compatible with the target scheduler, enabling their integration into the deployment planning process.

%Energy Mix vs Carbon Intensity

%Constraints Library: a set of modules, each one describing a specific constraint and how to implement it (which data to monitor, how to set thresholds, ... ). Constraints can be extended, users can define additional constraints and include them in the library (not only energy). 
\subsection{Application and Infrastructure Model}\label{ssec:input}

As shown in Figure~\ref{fig:architecture}, the \texttt{Green-aware Constraints Generator} uses several inputs for the generation of the deployment constraints. While some of them are standard (i.e., the application deployment plan, the monitoring metrics, and the grid carbon intensity) and do not require much explanation, here we focus on the two main artefacts:
\begin{itemize}
     \item \textbf{Application Description:} a specification of the application $\mathcal{A}$, including the set of services, and their inter-dependencies and characteristics;
    \item \textbf{Infrastructure Description:} a representation of the target infrastructure $\mathcal{I}$, comprising the available nodes where the application services may be deployed and the nodes' capabilities.
\end{itemize}

\paragraph{Application Description}\label{ssec:application}

The approach proposed in this paper aims to simplify and enrich the deployment planning process for microservice-based applications. An application is modelled as a set of cooperating services, each of which is independently deployable. To enable effective deployment, the application description must include relevant metadata that captures both functional and non-functional properties for each service.
Formally, the application description is denoted as $\mathcal{A}$, and the following properties characterise each service within it is:
\begin{itemize}
    \item \texttt{componentID}: a unique identifier for the service;
    \item \texttt{description}: a textual explanation of the functionalities provided by the service;
    \item \texttt{mustDeploy}: a flag indicating whether the service is mandatory or optional. Optional services may be excluded from the deployment plan if constraints, such as budget limitations, prevent their inclusion;
    \item \texttt{flavours}: set of available versions in which the service’s functionality can be implemented. The presence of multiple flavours allows greater flexibility in composing the application and facilitates adaptation to varying execution contexts;
    \item \texttt{flavoursOrder}: a preference list over the available flavours, indicating the order in which the application developer would prefer them to be selected. Higher-priority flavours are chosen whenever the deployment budget allows.
\end{itemize}

The inclusion of flavours and optional services in the application description aligns with best practices for sustainable software design, as discussed in~\cite{vitali2022towards}. Each flavour may differ in terms of computational demands and energy consumption. The flexibility to select among flavours and omit non-essential services enhances the sustainability of the application.

Complementing the application description, the requirements specification, denoted as $\mathcal{R}$, captures both functional and non-functional constraints for services and their flavours. Specifically:
\begin{itemize}
    \item \textbf{flavour level requirements} specify the computational resources needed for execution (e.g., CPU, RAM, Storage), as well as Quality of Service (QoS) constraints (e.g., availability);
    \item \textbf{service level requirements} include flavour-independent constraints such as security needs (e.g., firewall, SSL support) and network placement (i.e., whether the service must be deployed in a private or public subnet);
    \item \textbf{communication level requirements} specify QoS constraints (e.g., availability and latency) of the communication between services. 
\end{itemize}

All the properties described above must be specified by the DevOps engineer and provided as input. To enable effective reasoning about energy-aware deployments, the description of each flavour is enriched with energy consumption estimates. These estimates include both the predicted energy usage of individual flavours and the energy overhead associated with data exchanges between interconnected services. To capture this information, an additional \texttt{energy} property is introduced. This property is automatically generated by the \textit{Energy Estimator} and is included for each associated flavour and each inter-service communication link. The values assigned to the \texttt{energy} property are derived from the analysis of available monitoring metrics.

\begin{comment}
\begin{figure}[t]
\begin{yamlcodep}{.47\textwidth}
\begin{lstlisting}[language=yaml]
title: Application Schema
type: object
properties:
  name:
    type: string
  components:
    type: object
    description: application components definitions
    additionalProperties:
      type: object
      properties:
        componentID:
          type: string
          description: Unique identifier
        must:
          type: boolean
        flavours:
          type: object
          description: Flavours of the component
          additionalProperties:
            type: object
            description: Flavour-specific configuration object
required:
  - name
  - components

definitions:
  component:
    type: object
    required: [componentID, must, flavours]
    properties:
      componentID: { type: string }
      must: { type: boolean }
      flavours: { type: object }

  flavour:
    type: object
\end{lstlisting}
\end{yamlcodep}
\end{figure}
\end{comment}

\paragraph{Infrastructure Description} \label{ssec:infrastructure}
In this work, we consider applications to be deployed across the cloud continuum, which is characterised by significant heterogeneity in both the capabilities and geographic locations of the infrastructure nodes. The infrastructure can include both public cloud, private cloud, and on-premise nodes. This heterogeneity also manifests in varying pricing models for computational resources. 
The infrastructure description is denoted as $\mathcal{I}$. Each node is described by the following properties:
\begin{itemize}
    \item \texttt{capabilities}: a set of attributes describing the node’s ability to fulfil the application services’ requirements. This includes computational and storage resources (e.g., CPU, RAM, disk), bandwidth (input/output), availability, security features (e.g., firewall, SSL, encryption), and the subnet to which the node is assigned (public or private);
    \item \texttt{profile}: general metadata about the node, including the \texttt{cost} associated with its resource usage and the estimated \texttt{carbon} intensity, expressed in grams of CO\textsubscript{2}-equivalent per kilowatt-hour of energy consumed (gCO\textsubscript{2}eq/kWh).
\end{itemize}

To support decisions that promote energy and environmental sustainability, the \textit{Energy Mix Gatherer} extends the node description with additional information about the node’s carbon intensity (\texttt{carbon}). This information can either be explicitly provided by the DevOps engineer when known (e.g., an edge node powered by solar energy) or automatically inferred from the node’s geographic location using publicly available regional carbon intensity data.

\section{Green Constraints Generation} \label{sec:constraints}

This section describes the green-aware constraint generation through the architecture’s modules.

%The process of constraint generation is presented in three steps. First, in Sect.~\ref{subsec:constraints_def}, we introduce the set of constraints designed to capture energy- and environment-related aspects of the deployment. Then, in Sect.~\ref{subsec:constraints_gen}, we describe how these constraints are automatically derived from historical monitoring data. Finally, in Sect.~\ref{subsec:constraints_exp}, we discuss how the generated constraints are enriched with explainability features, enabling DevOps engineers to understand their rationale and potential impact on the deployment strategy.

\subsection{Energy Estimator}\label{ssec:monitoring}

The proposed approaches enrich the application description by leveraging historical monitoring data. These data are derived from past executions of the application and provide valuable insights. Focusing on the energy perspective, we collect and analyse information related to two key aspects:

\begin{itemize}
    \item \textbf{Computation energy profile:} For each service and its respective flavour, we collect data on the energy consumed during its execution. These data are available only if the service has previously been deployed with that flavour; otherwise, an estimation must be inferred. The average energy consumption for a service $s$ in flavour $f$ is computed as:
    \begin{equation}
        \mathit{energyProfile}(s,f) = \frac{1}{T} \sum_t \mathit{energy}_t(s,f)
        \label{eq:computationProfile}
    \end{equation}
    \noindent where $\mathit{energy}_t(s,f)$ denotes the energy consumed by service $s$ in flavour $f$ at time $t$, and $T$ is the number of available measurements.

    \item \textbf{Communication energy profile:} For each service and flavour, we also consider the energy consumption resulting from communication with other services. The communication energy profile from service $s$ (in flavour $f$) to service $z$ is computed as:
    \begin{equation}
        \mathit{energyProfile}(s, f, z) = \frac{1}{T} \sum_t \mathit{energy}_t(s, f, z)
         \label{eq:communicationProfile}
    \end{equation}
    \noindent where $\mathit{energy}_t(s, f, z)$ represents the energy consumed for data exchange from service $s$ in flavour $f$ to service $z$ at time $t$, and $T$ is the number of observations. We assume that the data transmitted by $s$ is not affected by the flavour of $z$ (services are independent and uncoupled), thus the energy cost of data transmission is independent of the flavour of the receiving service.
\end{itemize}

It is worth noting that the energy profile provides a hardware-agnostic, statistical estimation derived from direct execution measurements on heterogeneous nodes. While not intended to be a fine-grained prediction, this choice avoids exhaustive per-node profiling, impractical in dynamic infrastructures, and enables flexible comparison of component energy costs independently of their eventual deployment. In future work, this estimation can be refined by introducing differentiated profiles for distinct classes of nodes (e.g., cloud or edge).

\subsection{Constraint Library} \label{subsec:constraints_def}

The \texttt{Constraint Library} defines the types of constraints that can be generated. This is implemented in a modular way, each module defining the way to evaluate, generate, and explain a specific type of constraint. In this paper, we focus on two primary types of constraints, related to the energy required to execute the individual application services and to the energy consumed during communication between services. However, the library can be extended to include additional types.

%\paragraph{Avoid Node Constraints}

The first category of constraints addresses the energy consumption and emissions generated by the deployment of application services in the infrastructural nodes available. 

\begin{definition}\label{def:avoid}
The \texttt{AvoidNode} constraint instructs the scheduler to avoid deploying a service $s$ in a specific flavour $f$ on a node $n$ if the deployment is associated with excessive energy consumption or a high environmental footprint.

The logic underpinning this constraint can be expressed using the following Prolog rule:
\begin{lstlisting}[language=Prolog]
suggested(avoidNode(d(s,f), n)) :-
    highConsumptionService(s, f, n).
\end{lstlisting}
where:
\begin{itemize}
    \item \texttt{suggested(avoidNode(d(s,f), n))} is the head of the rule, indicating a recommendation to avoid deploying service $s$ (in flavour $f$) on node $n$;
    %\item \texttt{deployedTo(c, f, \_)} is a predicate that confirms component $c$ in flavour $f$ is being considered for deployment, regardless of the specific target node;
    \item \texttt{highConsumptionService(s, f, n)} is a predicate identifying deployments that lead to high energy usage or emissions. %The variable $W$ represents the quantified impact, which can subsequently inform ranking or filtering decisions.
\end{itemize}
\end{definition}

The second category of constraints addresses the energy consumption caused by data exchange between interacting services of the application. Communication between services deployed on different nodes can generate significant energy costs, particularly when large volumes of data are involved in their interaction. To mitigate this, the \texttt{Constraint Generator} can suggest co-locating such services to reduce the energy footprint associated with inter-node communication.

\begin{definition}\label{def:affinity}
The \texttt{Affinity} constraint guides the scheduler to deploy a service $s$ in a specific flavour $f$ on the same node as another service $z$ when their interaction involves exchanging large volumes of data and would otherwise result in high communication energy consumption.

The logic underpinning this constraint can be expressed using the following Prolog rule:
\begin{lstlisting}[language=Prolog]
suggested(affinity(d(s,f), d(z,_))) :-
    dif(s, z),
    highConsumptionConnection(s, f, z).
\end{lstlisting}
where:
\begin{itemize}
    \item \texttt{suggested(affinity(d(s,f), d(z,\_)))} is the head of the rule, expressing a recommendation to co-locate service $s$ (flavour $f$) and service $z$ (independently on its flavour) due to the associated environmental impact if they are deployed on separate nodes;
    %\item \texttt{deployedTo(c, fc, n)} and \texttt{deployedTo(s, fs, m)} specify the selection of the two components to be deployed;
    \item \texttt{dif(s, z)} % and \texttt{dif(n, m)}
    ensures that the rule applies only when $s$ and $z$ are two distinct services;
    \item \texttt{highConsumptionConnection(s, f, z)} is a predicate identifying that the communication between service $s$ in flavour $f$ and service $z$ generate a high energy consumption.
\end{itemize}
\end{definition}

\subsection{Constraint Generator} \label{subsec:constraints_gen}

Based on the types of constraints defined in the \texttt{Constraint Library}, the \texttt{Constraint Generator} derives green-aware deployment constraints from the analysis of monitored data. Each generated constraint~$c$ is assigned an estimated impact $Em$, in terms of environmental footprint, that satisfying the constraint is expected to achieve. This dynamic, data-driven approach avoids the limitations of static constraint definitions, which may not accurately reflect the real-time energy and environmental footprint of the application. In this section, we describe the process by which the \texttt{Constraint Generator} generates these constraints and computes their corresponding impact.

The \texttt{AvoidNode} constraints are derived from the historical energy consumption data for each flavour and the energy mix of the nodes. This information is made available by the \texttt{Energy Estimator} and the \texttt{Energy Mix Gatherer}, enriching the \texttt{Application Description} and the \texttt{Infrastructure Description} (Sect.~\ref{ssec:input}). Using the input data, the predicate \texttt{highConsumptionService(s, f, n)} is evaluated for every potential combination of service $s$, flavour $f$, and node $n$. The service and the node must have compatible network placement requirements (e.g., a private service can't be deployed in a public node). The predicate holds when the product of a flavour’s energy consumption and a node’s carbon intensity exceeds a predefined threshold $\tau$. Formally:
\begin{equation}
\begin{aligned}
    &\texttt{highConsumptionService}(s, f, n) \\ 
    &\text{if} \quad energyProfile(s,f) \cdot carbon(n) > \tau
    \end{aligned}
\end{equation}

For each valid combination satisfying the predicate, an \texttt{AvoidNode} constraint is instantiated. %Each constraint is associated with a weight $W$ representing its relative importance with respect to other constraints. This weight reflects the environmental cost of deploying service $s$ in flavour $f$ on node $n$, normalized against the worst-case environmental impact observed across the infrastructure and application:
% \begin{equation}
%     W = \frac{energyProfile(s,f) \cdot carbon(n)}{\max(energyProfile) \cdot \max(carbon)}
% \end{equation}

%\noindent where $max(carbon)$ is the highest carbon intensity available across all nodes, and $max(energyProfile)$ is the maximum energy consumption observed among all service flavours and communications. This normalization ensures that weights are bounded within the $[0,1]$ range and can be directly compared or prioritized.

The \texttt{Affinity} constraints are derived from the historical energy consumption for the communication between services. Using the input data, the predicate \texttt{highConsumptionConnection(s, f, z)} is evaluated for every potential combination of services $s$, flavour $f$, and receiving service $z$. The predicate holds when the communication energy exceeds a predefined threshold $\tau$:
\begin{equation}
\begin{aligned}
    &\texttt{highConsumptionConnection}(s, f, z) \\ 
    &\text{if} \quad energyProfile(s,f,z) > \tau
    \end{aligned}
\end{equation}

For each valid combination satisfying the predicate, a corresponding \texttt{Affinity} constraint is instantiated. %Each constraint is associated with a weight $W$ representing its relative importance with respect to other constraints: 
% \begin{equation}
%     W = \frac{energyProfile(s,f,z)}{max(energyProfile)} 
% \end{equation}

%\noindent In this case, since the deployment nodes for the two interacting services are not known in advance, it is not possible to accurately estimate the environmental impact based on the nodes' carbon intensity. Therefore, the weight $W$ associated with each \texttt{Affinity} constraint is computed solely based on the energy consumption related to the data exchange between services, omitting the carbon emission factor. As with the \texttt{AvoidNode} constraints, the weights for \texttt{Affinity} constraints are normalized within the $[0,1]$ range. This normalization allows for a direct comparison of the relevance of \texttt{Affinity} constraints against \texttt{AvoidNode} constraints.

The threshold $\tau$ is introduced to reduce the number of constraints and to avoid generating constraints that may lead to infeasible solutions for the solver. Choosing an appropriate value for $\tau$ is non-trivial: a high value may exclude relevant constraints that could improve the environmental impact of the deployment plan, whereas a low value may produce too many constraints, shrinking the scheduler’s search space and potentially yielding infeasible solutions. To address this, we adopt an adaptive strategy in which $\tau$ is computed from the distribution of the expected environmental impact of all services and communications observed in the monitoring history. Following the Pareto Principle~\cite{box1986analysis}, we assume that most of the potential energy savings come from a relatively small subset of constraints. According to this: 
\begin{equation}\label{eq:tau}
    \tau = q_{\alpha}, \quad \text{with } q_\alpha = \inf \{ x \;|\; F(x) \geq \alpha \}
\end{equation}
where $F(x)$ is the cumulative distribution function of the observed impacts. In the experiments, we set $\tau = q_{0.8}$, meaning that only the 20\% most impactful constraints are retained. We discuss the effect of $\tau$ on the constraints generation in Sect.~\ref{subsection:thscalability}.

\subsection{Knowledge Base and KB Enricher}\label{subsec:kb}

Constraint generation at each deployment should rely not only on recent observations but also on previously learned rules and monitored data. To this end, the process is supported by a \texttt{Knowledge Base}:
\begin{equation}
    KB = \langle SK, IK, NK, CK \rangle
\end{equation}

$SK$ stores knowledge about the energy behaviour of application services. Since the same service may exhibit different profiles across time (e.g., when deployed on different nodes or flavours), the KB aggregates their environmental footprint as:
\begin{equation}
    \langle s,f \rangle_t \;\mapsto\; \langle Em_{max}, Em_{min}, Em_{avg} \rangle
\end{equation}
\noindent where $s$ is a service, $f$ its flavour, and $t$ the last update time. The tuple captures the maximum, minimum, and average footprint observed from monitoring data.

$IK$ describes the environmental profile of data exchanges between services. For each communication, the KB stores:
\begin{equation}
    \langle s,f,z \rangle_t \;\mapsto\; \langle Em_{max}, Em_{min}, Em_{avg} \rangle
\end{equation}
\noindent where $s$ is the source service, $f$ its flavour, $z$ the destination service, and $t$ the last update time. The tuple captures the maximum, minimum, and average footprint observed from monitoring data.

$NK$ describes the environmental profile of infrastructure nodes. As node emissions vary over time, the KB stores for each node:
\begin{equation}
    n_t \;\mapsto\; \langle CI_{max}, CI_{min}, CI_{avg} \rangle
\end{equation}
\noindent where $n$ is a node, $t$ is the last update time, and the tuple denotes its maximum, minimum, and average carbon intensity.

Finally, $CK$ stores constraints learned in previous iterations. For each constraint:
\begin{equation}
    c_t \;\mapsto\; \langle Em, \mu \rangle
\end{equation}
\noindent where $c$ is the constraint (as defined in Section~\ref{subsec:constraints_def}), $Em$ the estimated footprint at generation time, and $t$ the generation timestamp. The weight $\mu$ is a \textit{memory weight} expressing constraint validity: if a constraint is not regenerated for several iterations, its reliability decreases accordingly.

The \texttt{KB Enricher} is responsible for keeping the KB up to date by integrating newly observed data and generated constraints, while adjusting the memory weight of previously learned but not re-observed constraints. In addition, it retrieves from the KB all valid past constraints to complement the new ones produced by the \texttt{Constraint Generator}. This ensures that, at each iteration, previously learned constraints with sufficiently high memory weight are properly considered in future deployment decisions. In the current implementation, the KB is realised as a semi-structured data store implemented through a collection of JSON files.

\subsection{Constraints Ranker} \label{subsec:ranking}

%Once relevant constraints are generated and retrieved from the KB, the \texttt{Constraints Ranker} assigns each of them an importance factor $w$:
% \begin{equation}
%      w = \frac{c_i.Em}{\max(c.Em) \in CK}
% \end{equation}
%\noindent where $c_i.Em$ is the estimated footprint associated to the constraint $c_i$ in the KB,$\max(c.Em)$ is the maximum footprint associated across all constraints in the KB. This normalization ensures that weights are bounded within the $[0,1]$ range and can be directly compared or prioritized.

Once the relevant constraints are generated and retrieved from the KB, the \texttt{Constraints Ranker} assigns each constraint $c_i$ a normalised importance weight $w_i$:  
\begin{equation}
     w_i = \frac{c_i.Em}{\max\limits_{c \in CK}(c.Em)}
\end{equation}

\noindent where $c_i.Em$ denotes the estimated footprint associated with constraint $c_i$ in the KB, and $\max_{c \in CK}(c.Em)$ is the maximum footprint among all constraints in the current knowledge base $CK$.  

This normalisation ensures that all weights $w_i$ fall within the interval $[0,1]$. As a result, constraints can be directly compared and ranked by their relative impact, with the most impactful constraint assigned a weight of 1. 

The importance weight $w_i$ can be used to guide the scheduler in prioritising among multiple constraints. However, even a constraint with high relative impact may have a negligible effect in terms of absolute emission savings. To account for this, the weights of constraints with low absolute impact are attenuated:
\begin{equation}
    w_i \leftarrow \lambda \, w_i
\end{equation}
where
\[
\lambda =
\begin{cases}
0.75, & \text{if } c_i.Em < F, \\
1,   & \text{otherwise,}
\end{cases}
\]
with \(F\) representing a minimum impact threshold. This ensures that constraints with minimal expected contribution have reduced influence on the deployment plan generation. At this stage, all constraints with $w_i < 0.1$ are discarded to avoid the generation of low-impact constraints.

\subsection{Explainability Generator} \label{subsec:constraints_exp}

After the constraint generation process, a set of constraints is produced, each annotated with an associated weight. These weights reflect the relative environmental benefit of satisfying each constraint, enabling the deployment decision process to balance constraint satisfaction with solution feasibility. Depending on the context, some constraints may be prioritised, while others may be relaxed if they conflict with more critical application requirements.

To support transparency and empower DevOps engineers with actionable insights, an additional artefact is generated alongside the constraint list: the \textit{Explainability Report}. For each constraint, the report includes a human-readable explanation detailing the rationale behind its creation. Furthermore, it provides an estimated range for the potential environmental gain, expressed as the minimum and maximum expected reduction in emissions if the constraint is enforced. This explainability layer supports informed decision-making, helping DevOps practitioners understand the trade-off involved and guiding them in the refinement of the final deployment plan.

\section{Validation} \label{sec:validation}

In this section, we present and discuss the experiments conducted to evaluate the effectiveness of the proposed methodology. We intentionally focus on the generation of sustainability-aware constraints and do not evaluate their integration into a scheduler for deployment plan generation. The design and feasibility of schedulers capable of handling such constraints have already been demonstrated in~\cite{gazza2025constraint, ponce2026failure} and therefore fall outside the scope of this paper. In particular,~\cite{ponce2026failure} evaluates the proposed energy constraint generation in both stressed simulation settings and realistic use cases, showing that the learned constraints effectively guide the scheduler in reducing environmental impact while satisfying performance and cost requirements. To highlight the adaptability of our approach and demonstrate how the generated constraints vary with changing contexts, we consider the following five scenarios:

\begin{itemize}
    \item \textbf{Scenario 1}: Serves as the baseline for our experiments. Starting from the application and infrastructure description provided in Section~\ref{subsection:casestudy}, we apply our methodology to generate energy-aware deployment constraints;   
    \item \textbf{Scenario 2}: Maintains the same application description as in Scenario 1 but replaces the infrastructure with a different set of nodes. This allows us to observe how the generated constraints adapt to changes in the environment;    
    \item \textbf{Scenario 3}: Simulates a variation in the carbon intensity of the infrastructure used in Scenario 1, reflecting real-world dynamics such as daily fluctuations in renewable energy availability. This scenario evaluates the responsiveness to changes in environmental conditions;   
    \item \textbf{Scenario 4}: Investigates the system's adaptability to application-level changes by simulating an update in service versions that alters the energy consumption profile of services. We assess how the generated constraints are modified to reflect the new application behaviour;   
    \item \textbf{Scenario 5}: Simulates an increased data exchange workload between services and tests the system's ability to adapt affinity constraints.
\end{itemize}

%We first provide additional details about the application and infrastructure descriptions used for the experimental evaluation in Sect.~\ref{subsection:casestudy}. We discuss the set of metrics and tools used to extract the energy-related monitoring information in Sect.~\ref{subsection:monitoringdataextraction}. We then present a comparison of the constraints generated across all scenarios in Sect.~\ref{subsection:constraintsgeneration}, and finally, we show the effectiveness of the \texttt{Green-aware Constraint Generator} in reducing the environmental impact in Sect.~\ref{subsection:proofofvalidity}.

The implementation of the \textit{Green-aware Constraint Generator}, along with all configuration files used in the experiments presented in this section, is publicly available in an open-source repository\footnote{https://github.com/POLIMIGreenISE/green-awareConstraintGenerator}.

\subsection{Case Study}
\label{subsection:casestudy}

To evaluate our proposed approach on a realistic multi-service application, we used Online Boutique\footnote{https://github.com/GoogleCloudPlatform/microservices-demo}, a benchmark developed by Google for testing microservice-based architectures. This web-based e-commerce demo consists of 10 microservices, which we extended and enriched with additional optional flavours (Table~\ref{table:services}), which can change the quality of the service, resulting in an impact on its energy consumption. In our setup, additional flavours are defined for the \textit{Frontend}, \textit{Checkout}, \textit{Recommendation}, and \textit{ProductCatalog} microservices, allowing us to analyse the energy-performance trade-offs involved in deployment decisions.
Some flavours indicate services with the same functionalities but less allocated resources, like the case for the \textit{Frontend} and \textit{Checkout}, whilst other flavours indicate services with reduced functionalities, like displaying fewer elements in a page, and that is the case for the \textit{Recommendation} and the \textit{ProductCatalog} services. %The energy profile associated to each flavour has been obtained by running the application on a Kuberenetes cluster. %The reported values are the average of XX executions.

% \begin{table}[t]
% \centering
% \begin{tabular}{l c c} 
%  \toprule
%  \textbf{Service} & \textbf{Flavour} & \textbf{Energy (kWh)} \\ 
%  \midrule
%  Frontend & Large & 1981 \\
%  Frontend & Medium & 1585 \\
%  Frontend & Tiny & 1189 \\
%  Checkout & Large & 134 \\
%  Checkout & Tiny & 107 \\
%  Recommendation & Large & 539 \\
%  Recommendation & Tiny & 431 \\
%  ProductCatalog & Large & 989 \\
%  ProductCatalog & Tiny & 791 \\
%  Ad & Tiny & 251 \\
%  Cart & Tiny & 546 \\
%  Shipping & Tiny & 98 \\
%  Currency & Tiny & 881 \\
%  Payment & Tiny & 34 \\
%  Email & Tiny & 50 \\
%  \bottomrule
% \end{tabular}
% \caption{Online Boutique: services and Energy Profiles}
% \label{table:services}
% \end{table}

\begin{table}[t]
\centering
\begin{tabular}{l c c} 
\toprule
\textbf{Service} & \textbf{Flavour} & \textbf{Energy (kWh)} \\ 
\midrule
\multirow{3}{*}{Frontend}       & Large  & 1981 \\
                                & Medium & 1585 \\
                                & Tiny   & 1189 \\
 \midrule           
\multirow{2}{*}{Checkout}       & Large  & 134 \\
                                & Tiny   & 107 \\
\midrule
\multirow{2}{*}{Recommendation} & Large  & 539 \\
                                & Tiny   & 431 \\
\midrule
\multirow{2}{*}{ProductCatalog} & Large  & 989 \\
                                & Tiny   & 791 \\
                                \midrule
Ad                              & Tiny   & 251 \\
\midrule
Cart                            & Tiny   & 546 \\
\midrule
Shipping                        & Tiny   & 98 \\
\midrule
Currency                        & Tiny   & 881 \\
\midrule
Payment                         & Tiny   & 34 \\
\midrule
Email                           & Tiny   & 50 \\
\bottomrule
\end{tabular}
\caption{Online Boutique: services and energy profiles (aggregated using multirow)}
\label{table:services}
\end{table}

To deploy the application, we consider two distinct groups of nodes, each associated with a specific carbon intensity\footnote{Carbon intensity values are derived using the Electricity Maps service: https://www.electricitymaps.com}. The first infrastructure configuration consists of five nodes located in various European countries (Table~\ref{table:nodesEU}), while the second configuration includes six nodes distributed across the United States (Table~\ref{table:nodesUS}). 
Since the carbon intensity significantly varies across regions, the emissions resulting from service execution do not depend only on the energy profile of the service itself but also on the carbon intensity of the hosting node. As a result, the placement of services across the available nodes becomes a crucial factor in minimising the application's overall environmental impact.

% \begin{table}[t]
% \centering
% \begin{tabular}{c c} 
%  \toprule
%     \textbf{Node} & \textbf{CI (gCO\textsubscript{2}eq/kWh)} \\
%  \midrule
%  France & 16 \\ 
%  \midrule
%  Spain & 88 \\
%  \midrule
%  Germany & 132 \\
%  \midrule
%  Great Britain & 213 \\
%  \midrule
%  Italy & 335 \\
%  \bottomrule
% \end{tabular}
% \caption{Infrastructure Description: Europe}
% \label{table:nodesEU}
% \end{table}

% \begin{table}[t]
% \centering
% \begin{tabular}{c c} 
%  \toprule
%     \textbf{Node} & \textbf{CI (gCO\textsubscript{2}eq/kWh)} \\
%  \midrule
%  Washington & 244 \\ 
%  \midrule
%  California & 235 \\
%  \midrule
%  Texas & 231 \\
%  \midrule
%  Florida & 570 \\
%  \midrule
%  New York & 236 \\
%  \midrule
%  Arizona & 229 \\
%  \bottomrule
% \end{tabular}
% \caption{Infrastructure Description: US}
% \label{table:nodesUS}
% \end{table}

% Europe Table
\begin{table*}[t]
\centering
\begin{tabular}{l c c c c c}
\toprule
\textbf{Node} & France & Spain & Germany & Great Britain & Italy \\
\midrule
\textbf{CI (gCO\textsubscript{2}eq/kWh)} & 16 & 88 & 132 & 213 & 335 \\
\bottomrule
\end{tabular}
\caption{Infrastructure Description: Europe}
\label{table:nodesEU}
\end{table*}

% US Table
\begin{table*}[t]
\centering
\begin{tabular}{l c c c c c c}
\toprule
\textbf{Node} & Washington & California & Texas & Florida & New York & Arizona \\
\midrule
\textbf{CI (gCO\textsubscript{2}eq/kWh)} & 244 & 235 & 231 & 570 & 236 & 229 \\
\bottomrule
\end{tabular}
\caption{Infrastructure Description: US}
\label{table:nodesUS}
\end{table*}

\subsection{Monitoring Data Extraction}
\label{subsection:monitoringdataextraction}

The demo application (Tab.~\ref{table:services}) was deployed on a Kubernetes cluster to collect a wide range of metrics. However, energy-related information is not natively available. In our approach, two main sources of energy consumption must be monitored: \textit{computation} and \textit{communication}, as introduced in Sect.~\ref{ssec:monitoring}. To this end, we integrated two monitoring tools, Istio\footnote{https://istio.io} and Kepler\footnote{https://sustainable-computing.io}, both exporting their metrics through Prometheus\footnote{https://prometheus.io}, a monitoring system for Kubernetes.
Kepler provides energy metrics for Kubernetes services and was used to monitor the consumption of individual services, enabling the derivation of the computation energy profile (Eq.~\ref{eq:computationProfile}). %Kepler reports values in Joules, which we converted into \verb|kWh|.
Istio was employed to monitor inter-service communications by capturing both the \textit{request volume} (requests per hour) and the \textit{request size} (GB). The data exchanged is directly proportional to communication energy consumption, which we estimated using the model in~\cite{aslan2018}:
\begin{equation}
    kWh = requestVolume \cdot requestSize \cdot k
\end{equation}
where $k$ denotes the transmission network electricity intensity (kWh/GB). This parameter is computed by dividing the total annual energy consumption of the internet by the total yearly data traffic. In this work, we use the projected value of $k$ for 2025, extrapolated from the trend reported in~\cite{aslan2018}. While acknowledging its limitations, we adopt this simple but widely used communication energy model due to the lack of observable, deterministic information about routing paths, geographic efficiency, and temporal network conditions at the application level.

%These metrics, once converted to the same scale, are able to provide a energy consumption estimate of the services and service connections, which will help us generating the final emissions of a given service or connection on any given node.

\subsection{Constraints Generation}
\label{subsection:constraintsgeneration}

%An energy-aware deployment is obtained through the adoption of energy-related constraints generated by the \textit{Energy Analyzer}. These constraints come in the form of \texttt{AvoidNode} constraints between a service in its flavour and a node or \texttt{Affinity} between two services.

In this section, we show the constraints generated by the \textit{Green-aware Constraint Generator} in the scenarios introduced before.

\textbf{Scenario 1} generates constraints for the deployment of the application described in Table~\ref{table:services} deployed on the infrastructure in Table~\ref{table:nodesEU}:
\begin{lstlisting}[language=Prolog]
affinity(d(frontend,large),
    d(productcatalog,large),0.01).
affinity(d(frontend,large),
    d(recommendation,large),0.002).
affinity(d(frontend,large),
    d(cart,tiny),0.003).
affinity(d(frontend,large),
    d(currency,tiny),0.007).
avoidNode(d(frontend,large),
    greatbritain,0.636).
avoidNode(d(frontend,large),italy,1.0).
avoidNode(d(productcatalog,large),
    italy,0.446).
\end{lstlisting}
As we can see, the \texttt{Affinity} constraints have a significantly lower weight than the \texttt{AvoidNode} constraints. This is because their estimated emissions are less relevant when compared to the energy generated by the computation of the services. The \texttt{Constraints Ranker} automatically removes them, leaving us with the final constraints composed of only \verb|avoid|.
\begin{lstlisting}[language=Prolog]
avoidNode(d(frontend,large),
    greatbritain,0.636).
avoidNode(d(frontend,large),italy,1.0).
avoidNode(d(productcatalog,large),
    italy,0.446).
\end{lstlisting}

In \textbf{Scenario 2}, we applied the approach considering a different infrastructure description, defined in Table~\ref{table:nodesUS}. We observe how the recommended constraints and their weights change.
\begin{lstlisting}[language=Prolog]
avoidNode(d(frontend,large),washington,0.428).
avoidNode(d(frontend,large),california,0.412).
avoidNode(d(frontend,large),florida,1.0).
avoidNode(d(frontend,large),newyork,0.414).
avoidNode(d(productcatalog,large), 
    florida,0.446).
\end{lstlisting}

The constraints generator will target the most consuming services (\textit{Frontend} and \textit{ProductCatalog}) to avoid their deployment on the less green nodes.

In \textbf{Scenario 3}, we analyse node carbon intensity degradation. Taking the same infrastructure used in \textbf{Scenario 1} (Table~\ref{table:nodesEU}), we simulate a degradation in the carbon intensity of the ``France'' node, becoming 376~gCO\textsubscript{2}eq/kWh, instead of the previous 16~gCO\textsubscript{2}eq/kWh. This can be due to the typical dynamicity of renewable energy sources. For the scenario at hand, we can assume that the node switched from a renewable source to a brown one. Looking at the generated constraints, we can see how weights are correctly changed to prioritise the avoidance of the France node for the most consuming services:
\begin{lstlisting}[language=Prolog]
avoidNode(d(frontend,large),france,1.0).
avoidNode(d(frontend,large),
    greatbritain,0.566).
avoidNode(d(frontend,large),italy,0.891).
avoidNode(d(productcatalog,large),
    france,0.446).
\end{lstlisting}

In \textbf{Scenario 4}, we analyse application-level changes, simulating a service becoming optimised and consuming less energy. The ``frontend'' has so far overshadowed the other services in terms of consumption. We simulate a new and more efficient version being released, reducing its energy consumption to 481 kWh. We can see how the outcome changes:

\begin{lstlisting}[language=Prolog]
avoidNode(d(productcatalog,large),italy,1.0).
avoidNode(d(currency,tiny),italy,0.89).
\end{lstlisting}

In \textbf{Scenario 5}, we analyse how the \textit{Energy Analyser} adapts when a higher data volume is exchanged between the application services. For our simulation, we assumed that the traffic volume could increase up to \textit{15'000} times compared to the normal monitored values (e.g., simulating video streaming instead of picture exchange). The resulting constraints are as follows:

\begin{lstlisting}[language=Prolog]
affinity(d(frontend,large),
    d(productcatalog,large),0.572).
affinity(d(frontend,large),
    d(currency,tiny),0.381).
avoid(d(frontend,large),greatbritain,0.636).
avoid(d(frontend,large),italy,1.0).
avoid(d(productcatalog,large),
    italy,0.446).
\end{lstlisting}

\noindent As can be observed, now the \texttt{Affinity} constraints become more relevant and are not filtered due to their increased weight. 

\subsection{Explainability Report}
\label{subsection:proofofvalidity}

Alongside each constraint, the \textit{Green-aware Constraint Generator} generates an \textit{Explainability Report}, which provides a human-readable explanation outlining the rationale behind each constraint and its potential environmental impact. In this section, we present the \textit{Explainability Reports} generated for \textbf{Scenario 1} as an illustrative example, highlighting the potential environmental benefits of adopting the suggested constraints.

%For \textbf{Scenario 1}, the following \textit{Explainability Report} is produced:

\begin{lstlisting}[language=Prolog]
An "AvoidNode" constraint was generated for the deployment of the "Frontend" service in the "large" flavour on the "GreatBritain" node. This decision was driven by the high resource consumption of the selected flavour combined with the poor energy mix of the target node.
The estimated emissions savings resulting from avoiding this deployment range between 390.38gCO2eq and 160.51gCO2eq.

An "AvoidNode" constraint was generated for the deployment of the "Frontend" service in the "large" flavour on the "Italy" node. This decision was driven by the high resource consumption of the selected flavour combined with the poor energy mix of the target node.
The estimated emissions savings resulting from avoiding this deployment range between 632.14gCO2eq and 241.76gCO2eq.

An "AvoidNode" constraint was generated for the deployment of the "ProductCatalog" service in the "large" flavour on the "Italy" node. This decision was driven by the high resource consumption of the selected flavour combined with the poor energy mix of the target node.
The estimated emissions savings resulting from avoiding this deployment range between 282.17gCO2eq and 107.91gCO2eq.
\end{lstlisting}

The proposed emission saving ranges are computed by comparing the avoided deployment with both the next worst (lower bound) and the optimal (upper bound) node choices, in terms of environmental efficiency. These ranges estimate how the service's footprint would change if the constraints are satisfied. While it is not feasible to always achieve the maximum possible savings, since doing so would unrealistically require deploying all services on the single most efficient node, respecting the constraints suggested by the methodology can still lead to a meaningful reduction in overall emissions.

\subsection{Scalability}
\label{subsection:scalability}

\begin{figure*}[t]
     \centering
     \begin{subfigure}[t]{0.95\textwidth}
         \centering
         \includegraphics[width=\textwidth]{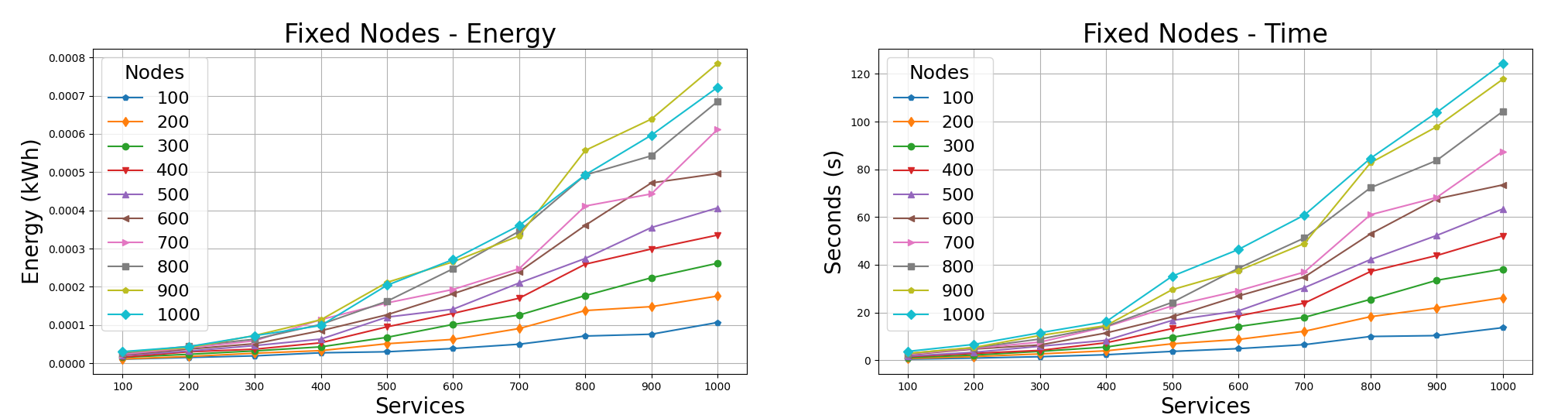}
         \caption{Application-level scalability}
         \label{fig:NodesFINAL}
     \end{subfigure}
     \begin{subfigure}[t]{0.95\textwidth}
         \centering
         \includegraphics[width=\textwidth]{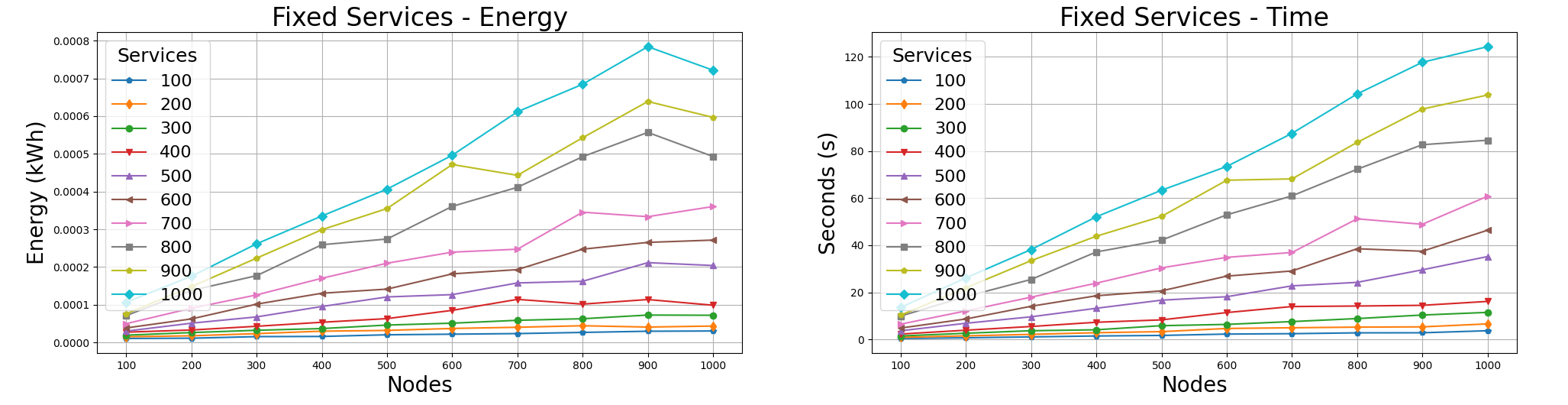}
         \caption{Infrastructure-level scalability}
         \label{fig:ServicesFINAL}
     \end{subfigure}
        \caption{Energy and time scalability for constraint generation}
        \label{fig:scalability}
\end{figure*}

The case study results demonstrate the ability of the proposed approach to automatically generate relevant deployment constraints in a realistic scenario. However, real-world applications and infrastructures can be significantly larger than the one considered in our experiments. In this section, we conduct a scalability study to evaluate the performance of our approach with applications containing more services and flavours, as well as infrastructures with a larger number of nodes.  
The scalability study measures the energy consumption and execution time required by the \textit{Green-aware Constraint Generator} to produce the constraints and the explainability report. These metrics were collected using the \textit{Code Carbon Emission Tracker}\footnote{https://codecarbon.io} and Python's native \textit{time} library.

Two types of experiments were conducted: 
(i) \textbf{Application-level scalability:} keeping the infrastructure fixed, we increased the number of application components; (ii) \textbf{Infrastructure-level scalability:} keeping the application configuration fixed, we increased the number of infrastructure nodes.
For application-level scalability, we ran 10 experiments with a fixed number of nodes, increasing the number of application components from 100 to 1,000 in steps of 100. Results are shown in Figs.~\ref{fig:NodesFINAL}. To account for variability, each data point represents the average of 10 executions. Results show that the energy consumption remains very low across all scenarios, making the effort to generate constraints reasonable relative to the environmental gains. Execution times are also reasonable, on the order of a few seconds, with a maximum of 120 seconds in the most extreme case. Given the low frequency at which constraint generation is required, these results indicate good scalability. Variability across runs is minimal; although slightly higher in configurations with larger numbers of nodes and components, it remains limited overall, confirming the robustness of the measurements.
Similar trends are observed for infrastructure-level scalability, as shown in Figs.~\ref{fig:ServicesFINAL}. Energy consumption and execution time grow approximately linearly with the number of nodes, remaining low in absolute terms. Comparing the two experiments, the number of application components appears to have the largest impact on both energy and time. % A combined view of application- and infrastructure-level scalability is presented in a 3D plot in Fig.~\ref{fig:3Dplot}. 
In the worst case, the constraint generation process has a computational complexity bounded by \(O(|S|\cdot|F|\cdot|N|)\); however, this upper bound is rarely reached in practice due to the optimisation mechanisms that avoid exhaustive enumeration of all combinations.

% \begin{figure*}[t]
%     \centering
%     \includegraphics[width=0.98\textwidth]{figures/finalGraphNodes.png}
%     \caption{Energy consumption for constraint generation (application-level scalability)}
%     \label{fig:NodesFINAL}
% \end{figure*}

% \begin{figure*}[t]
%     \centering
%     \includegraphics[width=0.98\textwidth]{figures/finalGraphServices.png}
%     \caption{Energy consumption for constraint generation (infrastructure-level scalability)}
%     \label{fig:ServicesFINAL}
% \end{figure*}

% \begin{figure*}[t]
%     \centering
%     \includegraphics[width=0.7\textwidth]{figures/EnergyAnalyzer3D.png}
%     \caption{3D plot of scalability experiments combining application- and infrastructure-level variations}
%     \label{fig:3Dplot}
% \end{figure*}

\subsection{Threshold Analysis}
\label{subsection:thscalability}

Section~\ref{subsec:constraints_gen} introduced the use of a threshold $\tau$ to limit the number of constraints generated by the methodology. This threshold plays a crucial role, as it directly influences both the feasibility of the scheduler and the environmental benefits of the resulting deployment plan. As formalised in Eq.~\ref{eq:tau}, $\tau$ is defined in terms of quantiles, ensuring that only constraints with sufficiently high expected impact are retained. 

In this section, we analyse how different values of $\tau$ affect the results, both in terms of the number of generated constraints and the expected environmental gains. The evaluation was conducted on a simulated scenario with 100 services and 100 nodes, each assigned randomised but realistic energy profiles.

Table~\ref{table:thresholds} reports the effect of varying $\tau$ across different quantile levels. As expected, lowering the quantile increases the number of generated constraints. This growth is not linear but accelerates rapidly, which can hinder the solver’s ability to find feasible solutions. These results highlight the importance of selecting an appropriate threshold to focus on the most impactful constraints, while avoiding an explosion in their number. 

\begin{table*}[t]
\centering
\begin{tabular}{l c c c c c c c c c} 
 \toprule
    \textbf{Quantile level} & 0.90 & 0.85 & 0.80 & 0.75 & 0.70 & 0.65 & 0.60 & 0.55 & 0.50\\
 \midrule
 \textbf{Constraints} & 85 & 137 & 227  & 371  & 636 & 804 & 1056 & 1164 & 1316 \\
 \bottomrule
\end{tabular}
\caption{Number of generated constraints as a function of the quantile threshold.}
\label{table:thresholds}
\end{table*}

\begin{figure*}[t]
    \centering
    \includegraphics[width=0.75\textwidth]{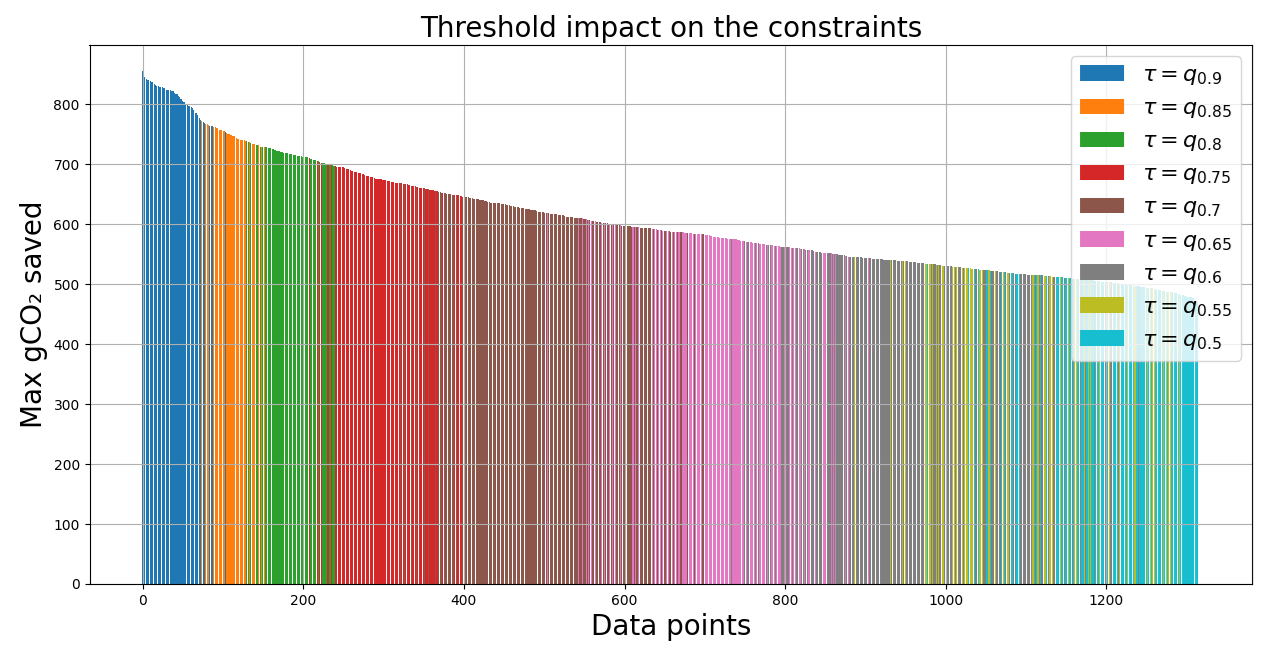}
    \caption{Distribution of potential emission savings across constraints for different quantile thresholds}
    \label{fig:Pareto}
\end{figure*}

To evaluate the effect of the threshold on potential environmental savings, we represented all generated constraints in Fig.~\ref{fig:Pareto}. Each constraint is shown as a bar, where the bar height corresponds to the potential emission reduction if the constraint is satisfied. Constraints are ordered from the most to the least impactful in terms of environmental footprint. Different colours indicate different quantile thresholds: each threshold includes the constraints of its own colour as well as those generated by stricter thresholds. As illustrated, the relative importance of constraints decreases when lowering the threshold, confirming that the most impactful constraints are concentrated at higher quantile levels. Considering the trade-off between the higher potential footprint savings of lower thresholds and the infeasibility caused by the large number of resulting constraints, selecting $\tau = q_{0.8}$ represents a balanced compromise.

\section{Conclusion} \label{sec:conclusion}

In this paper, we introduced an approach for the automatic and adaptive generation of sustainability-aware deployment constraints for microservice-based applications in cloud–edge infrastructures. These constraints are derived from the analysis of runtime monitoring data, taking into account the energy consumption of application components, their communication intensity, and the sustainability characteristics of infrastructure nodes. The ultimate objective is to support the generation of deployment plans enriched with context-aware recommendations that foster environmentally efficient decisions. To enhance transparency and usability, particularly for DevOps engineers, our approach complements the actionable constraints with an \textit{Explainability Report}, which clarifies the rationale behind each recommendation and quantifies its potential environmental benefits.

We validated the methodology on a benchmark microservice application under multiple dynamic scenarios, including variations in infrastructure, application behaviour, and energy contexts. The experiments confirmed the adaptability of the approach and demonstrated its ability to reduce the application’s carbon footprint through informed constraint generation. Furthermore, the scalability study showed that the proposed solution remains efficient even for large infrastructures and complex applications, both in terms of energy consumption and execution time.

As future work, we plan to broaden the set of supported constraints to include scenarios with batch-processing components. In addition, we aim to deepen the integration of the proposed constraint generation mechanism with an existing scheduler, capable of producing deployment plans directly from the generated constraints, to better evaluate the actual environmental impact of our approach in real deployment settings.

\section*{Declaration on the use of generative AI}

During the preparation of this work, the authors used ChatGPT-4 in order to check and improve the writing. After using this tool, the authors reviewed and edited the content as needed and take full responsibility for the content of the published article.

\section*{Acknowledgment}
This work was supported by the FREEDA project (CUP:
I53D23003550006), funded by the frameworks PRIN (MUR, Italy) and by the European Union (TEADAL, 101070186).

\balance
\bibliographystyle{elsarticle-num} 
\bibliography{references}
\end{document}